\documentstyle[aps,epsfig]{revtex}
\newcommand{\om}{\Omega_{\rm M}}
\newcommand{\ola}{\Omega_\Lambda}

\begin{document}
\twocolumn
\title{Gravitational lensing of the farthest known supernova SN1997ff}
\author{
  E. M\"ortsell\thanks{E-mail address: edvard@physto.se},
  C. Gunnarsson\thanks{E-mail address: cg@physto.se}, and
  A. Goobar\thanks{E-mail address: ariel@physto.se}
}
\address{
  Department of Physics, AlbaNova, Stockholm University, \\
  SE--106 91 Stockholm, Sweden
}
\maketitle
 
\begin{abstract}
We investigate the effects of gravitational lensing due to intervening
galaxies on the recently discovered Type Ia supernova at $z\sim 1.7$, 
SN1997ff, in the 
Hubble Deep Field North. We find that it is possible to obtain a wide range of 
magnifications by varying the mass and/or the velocity dispersion 
normalization of the lensing galaxies. 
In order to be able to use SN1997ff to constrain the redshift-distance
relation, very detailed modeling of the 
galaxies to control the systematic effects from lensing is necessary.
Thus we argue, that based on our current limited knowledge of the 
lensing galaxies, it is difficult to use SN1997ff to 
constrain the values of $\om$ and $\ola$, or even to place severe limits
on grey dust obscuration or luminosity evolution of Type Ia supernovae. 

\end{abstract}
\vspace{3mm}


\section{Introduction}\label{sec:intro}
A major goal of cosmology is to determine the values of various cosmological
parameters such as the energy density in pressure-less matter, $\om$, and the
energy density in some energy component with negative pressure, e.g., the cosmological 
constant $\ola$.
It has been long recognized that a possible method to accomplish this goal is 
to constrain the redshift-distance relation through the study of
Type Ia supernovae (SNe). Since the redshift-distance relation is more sensitive
to different values of the cosmological parameters at high redshifts, the
recently discovered SN at $z\sim 1.7$ \cite{riess2001} 
might prove to be invaluable in this respect.

However, different systematic effects such as obscuration by grey dust, 
luminosity evolution of Type Ia SNe and gravitational lensing are also possibly more 
severe at high redshifts. In Ref.~\cite{point,ramme}, 
the systematic effects of gravitational lensing 
on a large sample of SNe have been investigated. 

In this paper, we investigate the effects of 
gravitational lensing due to galaxies lying close to the line-of-sight to
SN1997ff, generalizing the work of Lewis and  Ibata \cite{lewis} and Riess
et al. \cite{riess2001} 
by investigating the combined
effects from a larger number of galaxies and estimating the masses and 
velocity dispersion of the lensing galaxies from the measured luminosities.  

Riess et al. \cite{riess2001} argued that the observed brightness of SN1997ff 
suggests that there cannot be a sizeable luminosity evolution for Type Ia's
nor significant extinction by dust. Our work shows that the possible lensing 
magnification effects are large enough that the data is also consistent with 
an intrinsically dimmer supernova, or with significant dust density 
along the line-of-sight.

\section{Method}\label{sec:method}
From the Hubble Deep Field North (HDF-N; \cite{url:hdfn}), 
we obtain the relative positions and redshifts of galaxies lying 
in the proximity of the line-of-sight to SN1997ff. 
When spectroscopic 
redshifts are available we will use these, otherwise we use the photometric 
redshifts. 
We will assume that the errors due to uncertainties in the positions and 
redshifts are negligible in comparison to the errors due to the 
uncertainties in the modeling of the lensing galaxies. 

\begin{figure}[t]
  \centerline{\hbox{\epsfig{figure=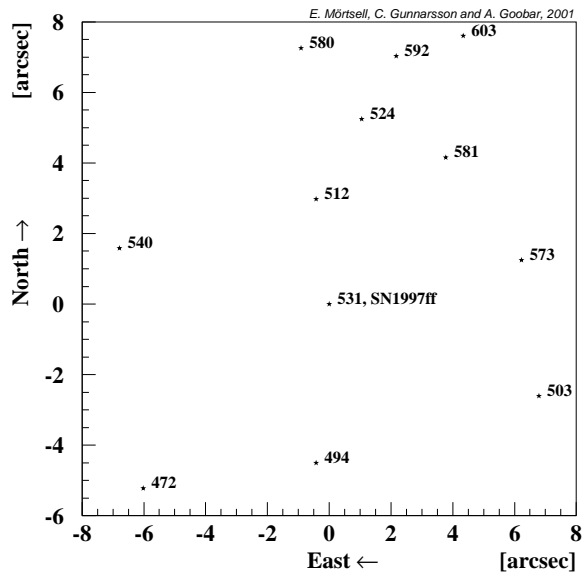,width=0.5\textwidth}}}
  \caption{The relative positions of galaxies lying closer than 10 arcseconds from 
    SN1997ff in the HDF-N. The corresponding redshifts, masses and velocity dispersions
    are tabulated in Table~\ref{tab:galaxies}.}
  \label{fig:hdfn} 
\end{figure}

In Fig.~\ref{fig:hdfn},
we plot the relative positions of the galaxies lying closer than 10 arcseconds 
to the line-of-sight
to SN1997ff  and its host galaxy No.~531.
We model the matter distribution of the galaxies as truncated isothermal
spheres with density profiles
\begin{equation}
  \label{eq:sis}
  \rho (r)=\frac{v^2}{2\pi}\frac{1}{r^2},
\end{equation}
where the velocity dispersions, $v$, are given by the Faber-Jackson
relation
\begin{equation}
  \label{FJ}
  \frac{v}{v_*}=\left(\frac{L}{L_*}\right)^{0.25}=10^{0.1(M-M_*)}.
\end{equation}
Here, $M$ is the absolute magnitude of the galaxy as measured in the $b_J$ 
magnitude system and $L$ is the luminosity. The star indicates typical galaxy values on $v$, $L$
and $M$ (and the mass, $m$, below).
We calculate $M$ by performing cross-filter K-corrections (assuming early-type galaxy spectra) 
on the observed magnitudes in
the $r$ and $i$ bands (ST magnitude system) into rest-frame $b_J$.

\begin{figure}[t]
  \centerline{\hbox{\epsfig{figure=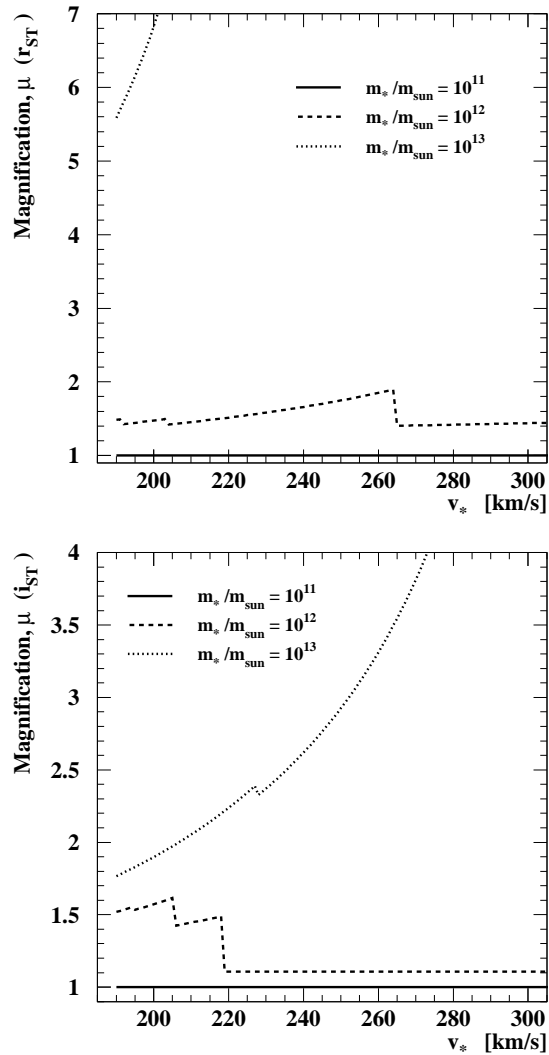,width=0.5\textwidth}}}
  \caption{The magnification, $\mu$, as a function of the typical galaxy velocity dispersion, 
    $v_*$, for three different values of the typical mass, $m_*$. In the upper panel,
    the velocity dispersion and masses of the galaxies are computed from the observed
    luminosities in the $r_{\rm ST}$-band, in the lower panel from the 
    $i_{\rm ST}$-band.}\label{fig:vdisvarri} 
\end{figure}

To estimate the masses of the lensing galaxies, we combine the observed
luminosities with the mass-to-luminosity ratio \cite{book:Peebles}
\begin{equation}
  \label{mtol}
  \frac{m}{m_*}=\left(\frac{L}{L_*}\right)^{1.25}=10^{0.5(M_*-M)},
\end{equation}
valid for early-type galaxies lying in the fundamental plane. 
We can derive an expression for the mass normalization, $m_*$, by assuming that
the luminosities of the galaxies are accurately described by the Schechter
luminosity function over the entire luminosity range, see \cite{bergstrom}  
\begin{equation}
  \label{eq:mrefestimate}
  m_{\ast}\sim 1.3\cdot 10^{14}\,\Omega_{\rm gal}\left(\frac{0.7}{h}\right)
  \left(\frac{10^{-2}h^3}{n_{\ast}\,{\rm Mpc}^3}\right)m_{\rm sun},
\end{equation}
where $\Omega_{\rm gal}$ is the mass density in galaxies and $n_*$ is the number density
of galaxies.
From the velocity dispersion and mass of the galaxies, we can compute
the cut-off radii for the projected mass (see below) of the halos as 
\begin{equation}
  \label{cut-off}
  d=\frac{m}{\pi v^2}.
\end{equation}
The galaxies are placed according to Fig.~\ref{fig:hdfn} at respective redshifts,
see Table~\ref{tab:galaxies}.
The magnification and deflection of the light from SN1997ff are calculated
by using the multiple lens-plane method
\cite[Ch.~9]{schneider}. The mass of each of the
lensing galaxies is projected onto a plane at the galaxy
redshift. The surface mass density obtained is, after being
appropriately scaled, called the convergence.
For a circularly symmetric lens, the convergence along with the total mass
within the radius of impact of the ray on the lens completely
characterizes the lensing effects.  
Next, the deflection angle and magnification in each plane is computed
by tracing light-rays backwards from the observer to the source, thereby
enabling us to determine the magnification and position of the ray. By
using a grid of
light rays, we are able also to determine the shape of an extended
source in the absence of lensing, i.e., its intrinsic shape, see
Sec.~\ref{sec:clues}. In Figs.~\ref{fig:vdisvarri} and
\ref{fig:mvarri}, SN1997ff is assumed to reside at exactly the host
galaxy center coordinates. 

Our calculations have been made using $\om =0.3$, $\ola =0.7$, $h=0.7$ 
and the filled beam approximation when 
calculating cosmological distances.
A typical value of $v_*=238$ km/s was obtained in Ref.~\cite{wilson} for
this cosmology.   

\section{Results}
\label{results}
In Fig.~\ref{fig:vdisvarri}, the magnification of SN1997ff is given as a function
of the normalization of the velocity dispersion of the galaxies for three different
values of the mass normalization, $m_*$. The velocity
dispersions are calculated from luminosities in the $r_{\rm ST}$-band (upper panel)
and $i_{\rm ST}$-band (lower panel).  
Note that the normalization of the velocity dispersion sets the concentration of the
galaxies [see Eq.~(\ref{eq:sis})] whereas the mass normalization sets the total mass 
of the galaxies.
For a circularly symmetric lens, the magnification is determined by the mass within 
the radius defined by the impact
parameter of the light-ray and the convergence at this radius, 
see Sec.~\ref{sec:method}. Since the size of the galaxies is inversely proportional
to the velocity dispersion squared, a higher velocity dispersion can mean a lower 
magnification since halos get smaller which may cause light-rays to pass outside halos, thus 
losing the convergence component in the magnification. It is this effect that causes the 
drops in the magnification curves in Fig.~\ref{fig:vdisvarri}.

\begin{figure}[t]
  \centerline{\hbox{\epsfig{figure=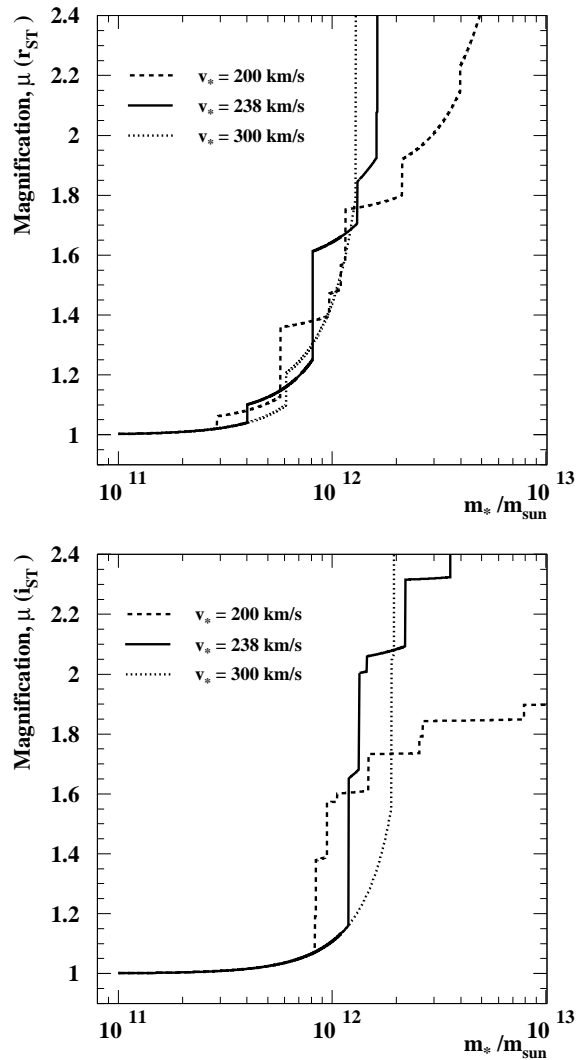,width=0.5\textwidth}}}
  \caption{The magnification, $\mu$, as a function of the mass normalization, 
    $m_*$, for three different values of the typical velocity dispersion, $v_*$. 
    In the upper panel,
    the velocity dispersion and masses of the galaxies are computed from the observed
    luminosities in the $r_{\rm ST}$-band, in the lower panel from the 
    $i_{\rm ST}$-band.}\label{fig:mvarri} 
\end{figure}

In Fig.~\ref{fig:mvarri}, the magnification is given as a function
of the normalization of the mass of the galaxies, $m_*$, 
calculated from luminosities in the $r_{\rm ST}$-band (upper panel)
and $i_{\rm ST}$-band (lower panel). 
Note that the magnification grows discontinously with $m_*$,
an effect similar to the drops in the magnification curves in 
Fig.~\ref{fig:vdisvarri}. In this case, it is due to the fact that larger 
masses gives larger halos, and thus it is possible to suddenly gain
the convergence component of the magnification
when the mass and accordingly the halo size are increased.
Of course, these effects are unphysical in the sense that they are very 
sensitive to the specific modeling of the halos, in this case the steepness of
the density profile and the cut-off radii. 
It is also an indication that lensing effects are very model dependent, 
and thus very detailed, individual modeling of the lensing galaxies is
necessary  
to make robust predictions of the magnification.


From our ray-tracing calculations, we find that in order to have multiple imaging
of SN1997ff, we need to have large values of $v_*$ and $m_*$. Also,
since there is no visible secondary image of the host galaxy, if multiply imaged,
the secondary image has to be very weak. We will therefore not treat the case of
multiple imaging of SN1997ff in this paper.

\newpage
\section{Clues from host galaxy}\label{sec:clues}
The appearance of the host galaxy might offer some clues to the magnitude of the 
lensing effect. The reason is that one expects an extended source to be tangentially
extended in the case of lensing. For an isothermal lens and a magnification factor
$\mu$, the image will be tangentially
stretched by a factor $\mu$ \cite{riess2001}. 
Since the host galaxy looks very close to
round (the minor to major axis ratio $b/a$ is 0.85), if highly magnified, the intrinsic shape 
of the galaxy will have to be very elliptical
and oriented in such a way as to counteract the tangential distortion due to lensing
effects. This effect has been used by Riess et al. \cite{riess2001} to argue that, 
assuming random intrinsic orientation and an intrinsic ellipticity distribution 
according to \cite{lambas}, the probability to observe such a small ellipticity in
the host galaxy is $\sim 14\%$ for a magnification of 0.4 mag and $\sim 3\%$ for a 
magnification of 0.8 mag. These probabilities were calculated for lensing from
galaxy No.~512 only.

\begin{figure}[t]
  \centerline{\hbox{\epsfig{figure=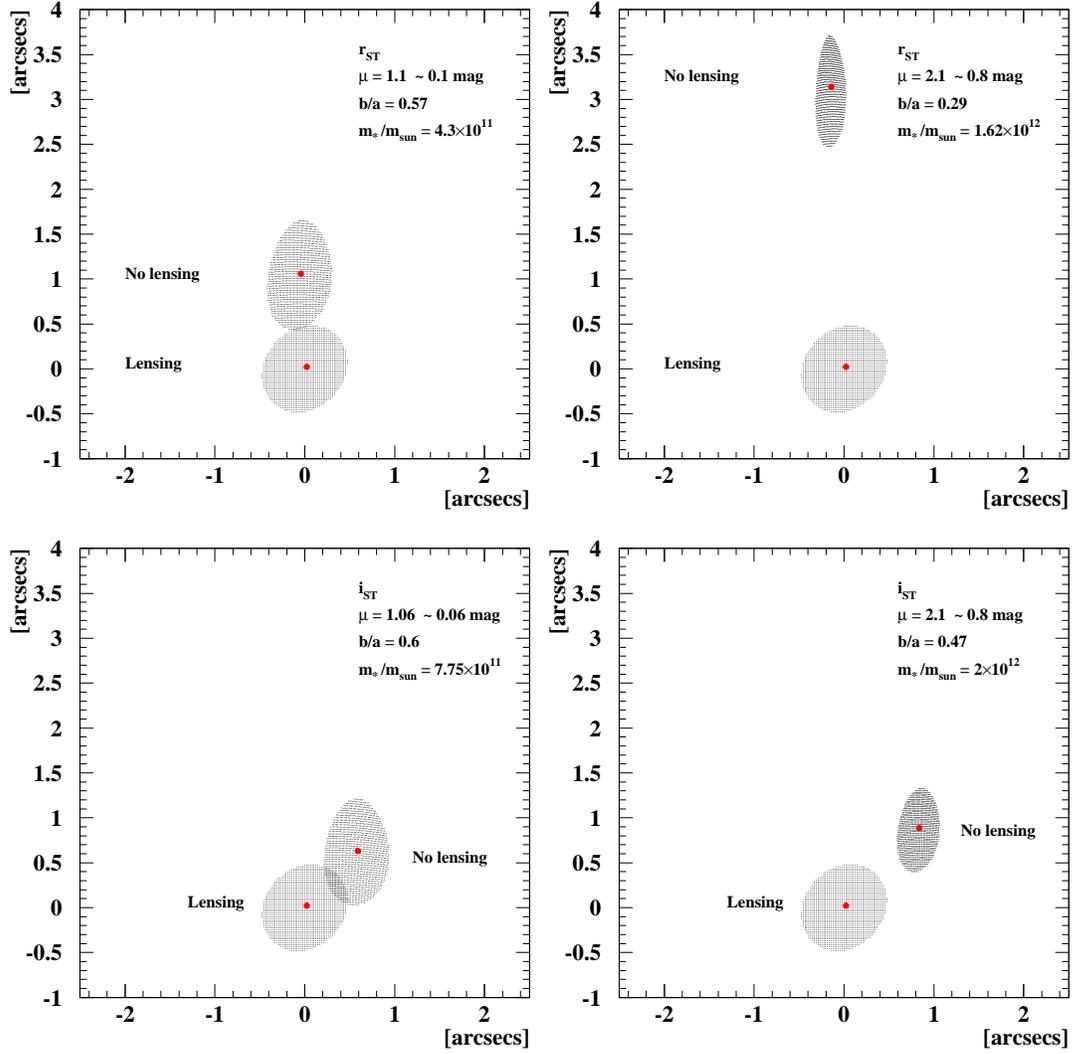,width=0.9\textwidth}}}
  \caption{The apparent (Lensing) and intrinsic (No lensing) shape of the 
    host galaxy No.~531, for different levels of magnification. The normalization
    of the velocity dispersion used is $v_*=238$ km/s. In the upper
    panel, masses and velocity dispersions are calculated from luminosities
    in the $r_{\rm ST}$-band, in the lower panel from the 
    $i_{\rm ST}$-band. The axis ratio of the image $b/a$ is 0.85.} \label{fig:elricomb} 
\end{figure}

In Fig.~\ref{fig:elricomb}, we have plotted the observed shape and orientation as 
well as the intrinsic shape and orientation of the
host galaxy for four different cases, 
including lensing effects from all nearby galaxies in the field. The
central point where we put the SN is indicated in a different shade.
In the upper panel, masses and velocity dispersions are calculated from luminosities
in the $r_{\rm ST}$-band, in the lower panel from the $i_{\rm ST}$-band.
We see that the intrinsic ellipticity of the host galaxy in fact only
has a rather mild
dependence on the magnification, $\mu$, in contrast to the analytical result for a
single isothermal lens. As an example, in the lower right panel of 
Fig.~\ref{fig:elricomb}, we have a magnification factor of $\mu=2.1$ (corresponding
to a magnification of 0.8 mag) with an intrinsic axis ratio of $b/a=0.47$ 
(corresponding to a E5 galaxy). From Fig.~2 in \cite{lambas} it is clear that
an axis ratio of $b/a=0.47$ is only a factor $\sim 1.5$ mor rare than
the observed ratio of $b/a=0.85$.  
Thus, comparing the results in Fig.~\ref{fig:elricomb} with the results of
\cite{riess2001} it is evident, that when including the effects from all galaxies
in the field, the constraints on the intrinsic ellipticity of the host galaxy is
considerably relaxed. Thus, we conclude that it is --- unfortunately --- hard to constrain
the magnification of SN1997ff from the apparent ellipticity of the host galaxy.

\section{Summary}\label{sec:summary}

We have investigated the effects of gravitational lensing on the magnification
of SN1997ff and the appearance of the host galaxy. Our results show that a large range 
of magnifications is possible for reasonable values of the galaxy masses and 
velocity dispersions. The value of the magnification is
very sensitive to details in the modeling of the matter distribution in the
lensing galaxies. Furthermore, we have found that the apparent (lack of)
ellipticity of the host galaxy does not put any strong constraints on the
magnitude of the magnification effect. 

Thus we conclude, that in order to use the apparent magnitude of a single high redshift 
SN to infer the values of $\om$ and $\ola$, or even to place meaningful limits 
on the possible dimming of Type Ia SNe by intergalactic grey dust or luminosity evolution, very careful modeling of the galaxies 
along the line-of-sight is needed in order to 
control the systematic effects from lensing. 

\section*{Acknowledgements}
The authors would like to thank Lars Bergstr\"om, Joakim Edsj\"o and Peter Nugent for helpful discussions and
Tomas Dahl\'en for providing us with galaxy spectral templates. 
AG is a Royal Swedish Academy 
Research Fellow supported by a 
grant from the Knut and Alice Wallenberg Foundation.
CG would like to thank the Swedish Research Council for financial support.

\begin{table}
  \caption{Data for galaxies in the HDF-N, closer than 10 arcseconds to
    the line-of-sight to SN1997ff. Redshifts in boldface are photometric redshifts
    used in cases where spectroscopic redshifts are not available.}
  \label{tab:galaxies}
    \begin{tabular}{cccccc} 
    No. & $z$ & $m/m_*(r_{\rm ST})$ & $m/m_*(i_{\rm ST})$ & 
    $v/v_*(r_{\rm ST})$ & $v/v_*(i_{\rm ST})$ \\ 
    \hline
    472 &  {\bf 1.107} &  2.863 &  0.700 &  1.234 &  0.931 \\
    494 &  0.873 &  2.481 &  0.670 &  1.199 &  0.923 \\
    503 &  0.847 &  0.675 &  3.197 &  0.924 &  1.262 \\
    512 &  0.555 &  0.986 &  0.520 &  0.997 &  0.877 \\
    524 &  0.557 &  0.802 &  1.106 &  0.957 &  1.020 \\
    540 &  {\bf 0.776} &  0.021 &  $8.1\cdot10^{-3}$ &  0.463 &  0.382 \\
    573 &  {\bf 0.468} &  $5.0\cdot10^{-4}$ & $5.3\cdot10^{-4}$ &  0.218 &  0.221 \\
    580 &  {\bf 1.500} & 19.378 &  2.185 &  1.809 &  1.169 \\
    581 &  {\bf 1.225} &  1.480 &  0.270 &  1.082 &  0.770 \\
    592 &  {\bf 0.776} &  0.194 &  0.062 &  0.721 &  0.573 \\
    603 &  {\bf 0.363} &  $1.4\cdot10^{-4}$ &  $4.7\cdot10^{-4}$ &  0.170 &  0.216 \\
    \end{tabular}
\end{table}

\onecolumn
\twocolumn


\vspace{-0.5cm}

\begin{references}

\bibitem{riess2001}
A.G.~Riess et al., 2001, {\it preprint} astro-ph/0104455


\bibitem{point}
E.~M\"ortsell, A.~Goobar and L.~Bergtsr\"om, {\it preprint} astro-ph/0103489

\bibitem{ramme}
R.~Amanullah et al., 2001, {\it in preparation} 

\bibitem{lewis}
G.F.~Lewis, and R.A.~Ibata, 2001, {\it preprint} astro-ph/0104254

\bibitem{url:hdfn}
{\tt http://astrowww.phys.uvic.ca/grads\\
/gwyn/pz/hdfn/spindex.html}

\bibitem{book:Peebles}
P.J.E. Peebles, {\em Principles of physical cosmology}
(Princeton University Press, Princeton, 1993).

\bibitem{bergstrom}
L.~Bergstr\"om, M.~Goliath, A.~Goobar, and E.~M\"ortsell, 
Astron.~Astrophys. 358 (2000) 13.

\bibitem{schneider} 
P.~Schneider, J.~Ehlers, and E.E.~Falco,
{\em Gravitational Lenses} (Springer-Verlag, Berlin, 1992)

\bibitem{wilson} 
G.~Wilson, N.~Kaiser, G.A.~Luppino and L.L.~Cowie,
{\it preprint} astro-ph/0008504
\bibitem{lambas}
D.G.~Lambas, S.J.~Maddox, and J.~Loveday, 1992, MNRAS, 258, 404
\end{references}
\end{document}